\documentclass{article}

\usepackage{PRIMEarxiv}

\usepackage[utf8]{inputenc} 
\usepackage[T1]{fontenc}    
\usepackage{url}            
\usepackage{booktabs}       
\usepackage{amsfonts}       
\usepackage{nicefrac}       
\usepackage{microtype}      
\usepackage{lipsum}
\usepackage[numbers]{natbib}
\usepackage{amsmath}
\usepackage{bbm}
\usepackage{setspace}
\usepackage{multicol}
\usepackage{multirow}
\usepackage{makecell}
\usepackage{fancyhdr}       
\usepackage{graphicx}       
\graphicspath{{media/}}     

\title{Bayesian modeling of dynamic behavioral change during an epidemic
}

\author{Caitlin Ward\textsuperscript{1,\thanks{Corresponding author: ward-c@umn.edu}}, Rob Deardon\textsuperscript{2, 3}, Alexandra M. Schmidt\textsuperscript{4}}

\begin{document}
\maketitle

\spacing{1.2}

\noindent \textsuperscript{1} Division of Biostatistics, University of Minnesota \\
\textsuperscript{2} Faculty of Veterinary Medicine, University of Calgary \\
\textsuperscript{3} Department of Mathematics and Statistics, University of Calgary \\
\textsuperscript{4} Department of Epidemiology, Biostatistics, and Occupational Health, McGill University

\vspace{10mm}

\begin{abstract}
For many infectious disease outbreaks, the at-risk population changes their behavior in response to the outbreak severity, causing the transmission dynamics to change in real-time. Behavioral change is often ignored in epidemic modeling efforts, making these models less useful than they could be. We address this by introducing a novel class of data-driven epidemic models which characterize and accurately estimate behavioral change. Our proposed model allows time-varying transmission to be captured by the level of “alarm” in the population, with alarm specified as a function of the past epidemic trajectory. We investigate the estimability of the population alarm across a wide range of scenarios, applying both parametric functions and non-parametric functions using splines and Gaussian processes. The model is set in the data-augmented Bayesian framework to allow estimation on partially observed epidemic data. The benefit and utility of the proposed approach is illustrated through applications to data from real epidemics.
\end{abstract}


\section{Introduction} \label{intro}

Human behavior is a driving factor in the spread of infectious disease through human populations. In the presence of increasing infection risk, individuals typically engage in protective behaviors to avoid becoming ill. These preventative behavior changes may be imposed by a governing body (e.g., city-wide lockdowns or school closures), or may be the result of personal choices (e.g., social distancing or voluntary masking). Behavioral change can have a substantial impact on the epidemic trajectory by delaying the peak, reducing the total number of individuals that contract the disease, and/or resulting in multiple waves of transmission. Additionally, behavioral change is dynamic; higher disease prevalence tends to result in increased preventative measures, which are subsequently relaxed as prevalence decreases. Understanding these changes in population behavior in response to an epidemic is crucial for public health practitioners and policy makers attempting to stop or slow the spread of the pathogen and allocate valuable resources.

Statistical modeling of infectious disease transmission provides a quantitative approach to understanding disease dynamics. The conventional methodology is based on the compartmental SIR model \cite{kermack1927}, which segments the population into \textbf{S}usceptible, \textbf{I}nfectious, and \textbf{R}emoved compartments capturing the important disease states. The model is then parameterized in terms of the rates of flow between compartments. Compartmental models can be implemented deterministically using ordinary differential equations or stochastically, with statistical inference typically carried out using Bayesian methodology. However, the traditional SIR model does not naturally account for transmission dynamics changing in real time as the population reacts to the outbreak, limiting its applicability to real epidemic data. 

Previous work incorporating behavioral change into the SIR model framework has been done in the deterministic setting, with models generally falling into one of three categories. One approach adds additional ``adherence" or ``awareness" compartments to capture reductions in susceptibility and transmissibility for individuals engaging in protective behaviors, with individuals more likely to be adhering to preventative measures when disease prevalence is high \cite{del2005effects, perra2011towards,  agaba2017mathematical, acuna2020modeling}. Another popular method uses a game-theoretic approach, supplementing the SIR model with a time-varying utility function balancing the costs associated with prevention strategies (monetary, liberty, social) with the benefit of lowering infection risk in the population \cite{reluga2010game, fenichel2011adaptive}. The remaining approach allows the transmission rate in the SIR model to be dynamically modified as a function of the recent disease trajectory \cite{capasso1978generalization, greenhalgh2015awareness, eksin2019systematic,  weitz2020awareness, franco2020feedback}. 

However, there are still several unanswered questions. First, can such deterministic approaches be translated to the stochastic Bayesian setting? The Bayesian paradigm offers the ability to impute partially observed epidemic data and incorporate prior knowledge about the disease process. However, Bayesian models often require simpler parameterizations than their deterministic counterparts to be computationally tractable \cite{andersson2012stochastic}. Many of the proposed models include highly complex depictions of behavioral dynamics, which would likely be challenging to implement in the Bayesian framework without detailed behavioral data. Second, can disease and behavioral change parameters be estimated from epidemic data? The aforementioned modeling efforts have focused almost entirely on model specification, with results coming from forward simulations using pre-specified parameter values. Notably absent in the literature is any assessment of the statistical properties of these models when fit to real data. Finally, how can these models be used in practice to increase understanding of human behavior during an epidemic? Behavioral change is anticipated during any real epidemic, but without any previous work fitting these models to data, their advantage is unknown.

In this article, we advance the field of stochastic infectious disease modeling by answering these important questions. We accomplish this by proposing a novel Bayesian SIR model formulation which captures dynamic behavioral change during an epidemic. The proposed model specifies the transmission rate as a function of recent disease occurrence, and computation is performed via Markov Chain Monte Carlo (MCMC) methods. In our simulation study, we thoroughly investigate the statistical properties of our Bayesian model when fit to data. In particular, we show that behavioral change parameters can be accurately estimated and that posterior predictions from the proposed model can detect subsequent peaks in incidence. To showcase the benefits of considering behavioral change in this class of models, we apply the model to data from an Ebola outbreak in the Democratic Republic of the Congo and the COVID-19 pandemic in New York City.

\section{Methods} \label{methods}

\subsection{Traditional SIR Model} \label{methods:traditional}

We model transmission using a discrete-time SIR model framework, where susceptible individuals can contract the infection from those who are infectious, and infectious individuals are removed when they no longer transmit the pathogen to others, due to death, isolation, or recovering with immunity. Let $t = 1, ..., \tau$ indicate discrete calendar time and $S_t$, $I_t$, and $R_t$ denote the number of individuals in the susceptible, infectious, and removed compartments in the continuous time interval $[t, t+1)$, respectively. Furthermore, define the transition vectors $I^*_{t}$ and $R^*_{t}$ to represent the number of individuals entering the indicated compartment in this interval. Compartment membership is temporally described by the set of difference equations: 
\begin{eqnarray} 
    S_{t+1} &=& S_t - I^*_t \label{diffEqs1} \\ 
    I_{t+1} &=& I_t + I^*_t - R^*_t \label{diffEqs2} \\ 
    R_{t+1} &=& R_t + R^*_t. \label{diffEqs3}
\end{eqnarray}
We assume a closed population, such that $S_t + I_t + R_t = N$ at all time points, where $N$ denotes the total population size. Given the population size, a set of initial conditions, and the transition vectors, the compartment membership vectors can be fully determined using Equations \ref{diffEqs1} - \ref{diffEqs3}.

In the Bayesian framework, we must establish the relationship between data and model parameters using probability distributions. We define the transitions between compartments to be binomially distributed  \cite{wilson1942epidemic, lekone}, such that $I^*_{t} \sim Bin\left(S_{t}, \pi_t^{(SI)} \right)$ and $R^*_{t} \sim Bin\left(I_{t}, \pi^{(IR)} \right)$. The transition probabilities $\pi_t^{(SI)}$ and $\pi^{(IR)}$ describe transmission of the pathogen and the duration of the infectious period, respectively. Assuming an independent Poisson contact process and constant probability of infection given a contact, the transmission probability, $\pi^{(SI)}$,  takes the form
\begin{equation} \label{eq:popAvgtransprob}
    \pi_t^{(SI)} = 1 - \exp \left(-\beta \frac{I_t}{N}\right).
\end{equation}
The parameter $\beta$ is interpreted as the transmission rate, which captures both the contact rate and the infection probability, as these are not separately identifiable \cite{brownReproductive}. The removal probability, $\pi^{(IR)}$, is derived by assuming the length of time an individual is infectious is exponentially distributed with rate $\gamma$. In discrete time, $\pi^{(IR)}$ is the conditional probability of transitioning on day $s + 1$, given the individual has remained infectious through time $s$, resulting in $\pi^{(IR)} = 1 - \exp \left(-\gamma\right)$. The parameter $\gamma$ is referred to as the removal rate, however, it is typically more interpretable to consider $1/\gamma$, the mean length of the infectious period.

The traditional SIR model assumes $\beta$ is constant over time, but this is generally not realistic. More likely, transmission changes as the population responds to the outbreak. Seasonal factors may also contribute, such as the start of the school year or changes in weather. This is incorporated by modifying the transmission probability from Equation \ref{eq:popAvgtransprob} as 
\begin{equation} \label{eq:pSIBetaT}
    \pi_t^{(SI)} = 1 - \exp \left(-\beta_t \frac{I_t}{N}\right),
\end{equation}
where $\beta_t$ is the transmission rate at time $t$. Changes in transmission can be modeled directly through covariates, such as change points corresponding to the timing of government intervention(s) \cite{lekone, ward2023individual} or measures of population mobility \cite{liu2020estimating, sartorius2021modelling}. This allows inference to be made on the relationship between covariate(s) and transmission. Covariates might not capture all important changes in transmission, so more flexible approaches have been proposed, including basis splines \cite{brownReproductive, hong2020estimation}, Gaussian processes \cite{xu2016bayesian}, or simple random walks \cite{irons2021estimating}. All of these approaches are limited when forecasting, as it is difficult to predict the lifting of a government lockdown, and restrictive assumptions must be made to allow any flexibly modeled trajectory of $\beta_t$ to continue into the future. Thus, an alternate approach accounting for the mechanism of behavioral change is needed.

\subsection{Behavioral Change (BC) Model} \label{methods:bcmodel}

The proposed behavioral change (BC) model allows for time-varying transmission which captures behavioral change via a dynamically structured dependence on previously observed epidemic trajectory. This is accomplished by allowing a constant transmission rate, $\beta$, to be modified by a time-varying level of alarm in the population, denoted $a_t$, such that $\beta_t = \beta (1 - a_t)$. We consider $a_t \in [0, 1]$, such that $a_t$ corresponds to the proportional reduction in transmission due to the alarm in the population. When $a_t = 0$, the population is in its natural `unalarmed' state, and transmission is described only by $\beta$. When $a_t = 1$, the population is in its maximal alarmed state and transmission is reduced to zero. Plugging this in to Equation \ref{eq:pSIBetaT} yields
\begin{equation} \label{eq:BCMtransprob}
    \pi_t^{(SI)} = 1 - \exp \left[-\beta  \left(1 - a_t \right) \frac{I_t}{N}\right].
\end{equation}
The structural dependence in the alarm is captured by specifying $a_t$ as a function of incidence smoothed over the past $m$ days, such that $a_t = f\left(\frac{1}{m} \sum_{i = t-m-1}^{t-1}I^*_i\right)$ and the smoothing parameter $m \in \{1, 2, ..., t-1\}$. For $t<m$, we use the moving average of the data up until time $t - 1$, e.g., at time $t = 3$ the alarm is based on the average of the incidence observed at times $t = 1 \text{ and } 2$. At the start of an epidemic, it is assumed that the alarm is zero. It is possible for the alarm function to depend on other reported metrics of epidemic severity, such as prevalence, hospitalizations, or test positivity rates, but we focus on an incidence-based alarm here. Incidence is advantageous for forecasting, as it is directly generated from the SIR model. Thus, incidence forecasts can be used directly to determine future alarm function values and generate further predictions. 

\begin{figure}[ht!]
\centering\includegraphics[width = 0.9\textwidth]{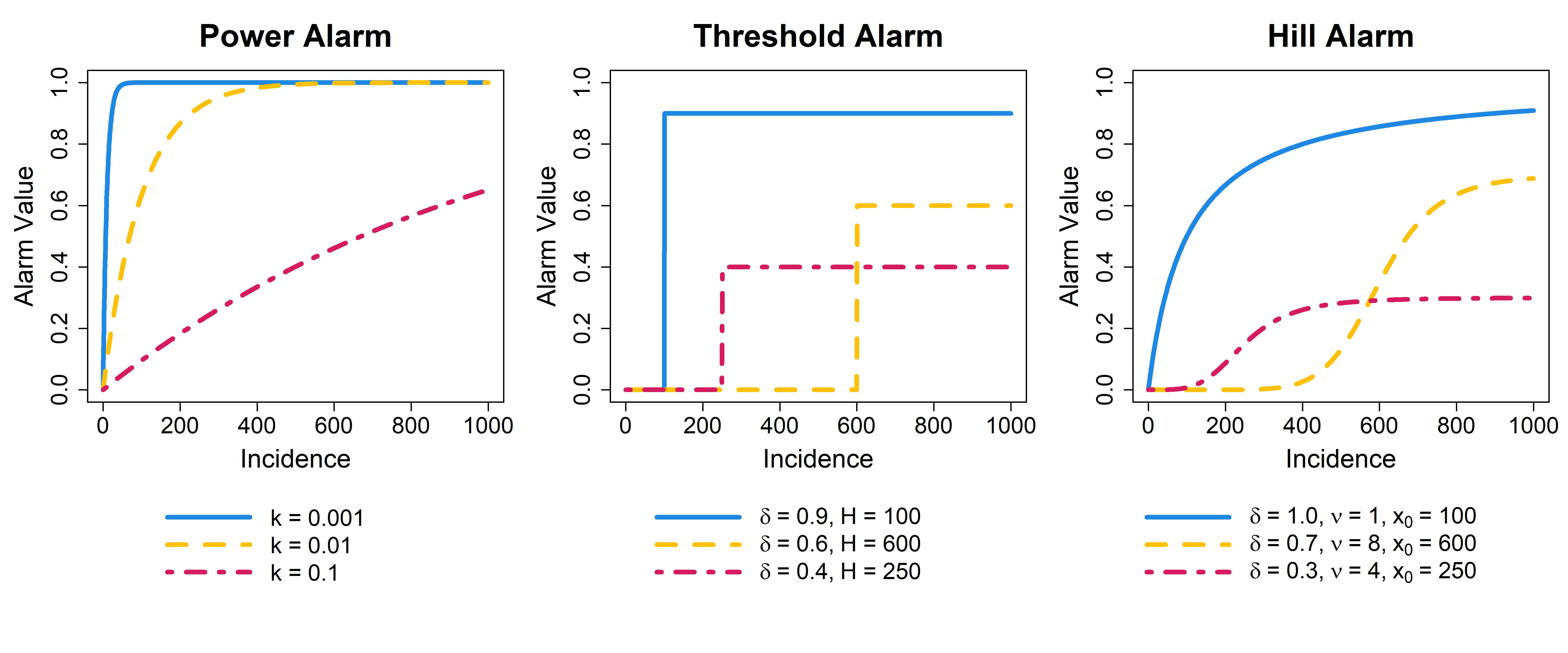}
\caption{Example alarm functions for various parameter specifications.}
\label{Fig1}
\end{figure}

Next, we must determine an appropriate functional form for the alarm function. As defined, the alarm must be between 0 and 1, and furthermore we would expect the alarm to be zero when there is no disease in the population, and to monotonically increase as the amount of disease in the population increases. There are many functions that satisfy these characteristics, and we investigate three possibilities of various complexity (Figure \ref{Fig1}). The first function considered is a one-parameter function $f(x) = 1 - (1 - x/N)^{1/k}$ used in previous deterministic literature \cite{eksin2019systematic, franco2020feedback}, which we call the ``power" alarm. The parameter $k > 0$ describes the growth rate, with smaller values corresponding with a faster rise in alarm. The next function we consider is a two-parameter constant change point model, which we call the ``threshold" alarm and specify as $f(x) = \delta \mathbbm{1}(x > H)$. The threshold alarm is zero until the threshold, $H$, is surpassed, at which point it becomes $\delta$. Note that $H$ can take on any value in the observed range of the data informing the alarm function and $\delta \in [0, 1]$. The final function analyzed is a modified Hill equation \cite{gesztelyi2012hill}, $f(x) = \frac{\delta}{1 + (x_0/x)^\nu}$, which we refer to as the ``Hill" alarm. With three parameters, the Hill alarm is the most complex of the three alarm functions considered; $\delta \in [0, 1]$ describes the asymptote, $x_0$ is the half occupation value, and $\nu > 0$ controls the growth rate. The Hill alarm can describe curves similar to the power alarm, as well as sigmoid-shaped curves that resemble a smoothed version of the threshold alarm. 

\begin{figure}[ht!]
\centering\includegraphics[width = 0.9\textwidth]{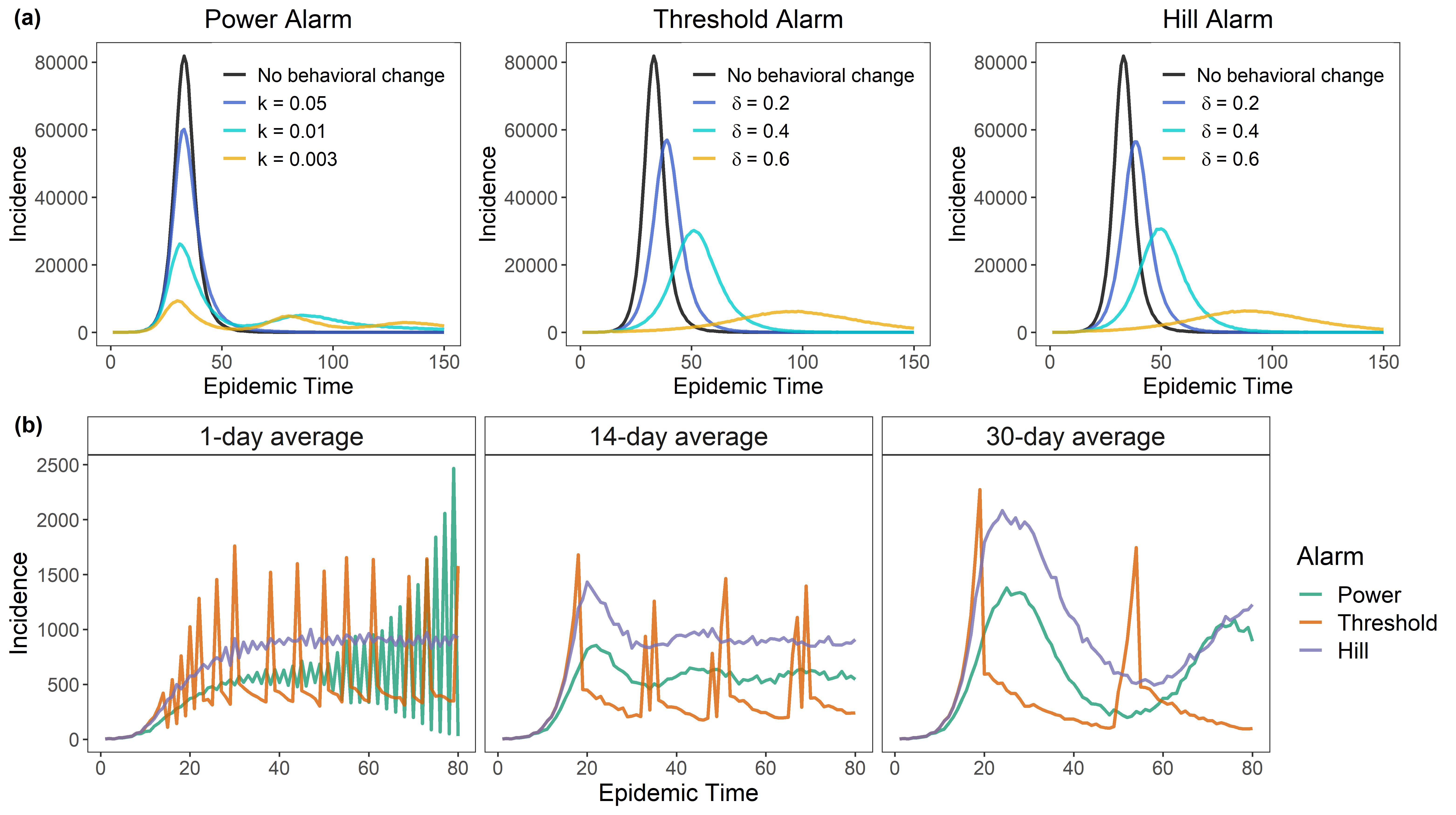}
\caption{Example simulated epidemics for various alarm function specifications.
(a) Compared to epidemics with no behavioral change, the power alarm with various $k$ values, 
the threshold alarm with $H = 100$ and various $\delta$ values, 
and the Hill alarm with $\nu = 5$, $x_0 = 100$, and various $\delta$ values. 
Across all simulations, $\beta = 0.6$ and $\gamma = 0.2$.
(b) Using 1-day, 14-day, and 30-day average incidence to inform the alarm function. 
Across the smoothing levels, the power alarm has $k$ = 0.0005, the threshold alarm has $\delta = 0.8$ and $H$ = 350, and the Hill alarm has $\delta = 0.85$, $\nu = 2$, and $x_0$ = 450. Across all simulations, $\beta = 0.6$ and $\gamma = 0.2$.}
\label{Fig2}
\end{figure}

Using these alarm functions in the BC model can generate a multitude of shapes of epidemic curves. Figure \ref{Fig2}a shows that higher levels of alarm reduce peak incidence. In addition, increasing the maximum alarm value in the threshold and Hill alarms delays the peak, while increasing the growth rate of the power alarm does not affect the timing of the initial peak. Figure \ref{Fig2}b illustrates epidemics generated using alarms which reach high values at relatively low levels of incidence, as well as how epidemic trajectory is affected by the amount of data informing the alarm function. When the alarm is based solely on the previous day's incidence, incidence becomes volatile and oscillates between levels producing high and low amounts of alarm. Peaks become more pronounced and spread out when the 14 or 30-day average incidence is used to inform the alarm. Finally, the threshold alarm, which reaches its maximum instantaneously, leads to epidemic curves with very sharp peaks. In contrast, the power and Hill alarm increase gradually, resulting in smoother peaks.

In practice, it may not be obvious which alarm function to use, or a method which does not restrict the shape of the alarm function may be preferred. For this reason, we also investigate the use of more flexible non-parametric approaches to estimating the alarm function using basis splines and Gaussian processes. The spline alarm is modeled using natural cubic splines with estimated knot locations and written as $f(x) = \mathbf{X_B}' \mathbf{b}$. The basis matrix, $\mathbf{X_B}$, is constructed across the range of observed $m$-day average incidence and $\mathbf{b}$ denotes the associated basis parameters. Constraints were used during estimation to ensure $f(x) \in [0, 1]$, but an appropriate link function (e.g., logit) could also be used. For the Gaussian process approach, we assume the logit of the alarm function is a realization from a multivariate normal distribution which is fully specified by its mean $m(x)$ and covariance $k(x, x')$, i.e., $\text{logit} \left[f(x)\right] \sim \text{MVN}\left[m(x), k(x, x')\right]$. We specify the mean function to start at $(0, 0)$ and end at $(\max (x), 1)$ on the logit scale, corresponding with the characteristics we expect of the alarm function. We generally expect the alarm function to be smooth and use the squared exponential covariance function $k(x, x') = \sigma^2 \exp\left[-(x - x')^2  / 2 l^2 \right]$, where $\sigma^2 > 0$ the signal variance controls scaling and $l > 0$ is the length-scale parameter controlling smoothness. Both alarms are defined across the range of observed $m$-day average incidence, so linear interpolation was used to find the value of the alarm on each day.

One of the most important quantities estimated by epidemic models is the reproductive number, denoted $\mathcal{R}_0$, which quantifies the spread of the pathogen in the population. Various methods of calculating $\mathcal{R}_0$ exist, and we use the approach of \cite{ward2023idd}. The effective reproductive number is calculated as $\mathcal{R}_0(t) = S_t \sum_{k = t}^{\infty} \left[1 - \exp(-\beta_k / N)\right] \exp(-\gamma)^{k - t}$ and provides the expected number of secondary infectious caused by a single individual that becomes infectious at time $t$ (for the full derivation, see \cite{ward2023idd}). $\mathcal{R}_0$ can be interpreted in relation to the threshold of 1, as $\mathcal{R}_0 \geq 1$ means the epidemic will continue to propagate through the population and $\mathcal{R}_0 < 1$ indicates the epidemic will eventually die out.

\subsection{Bayesian Estimation and Implementation} \label{methods:estimation}

The complete log-likelihood for the chain binomial SIR model is
\begin{equation}
\begin{split}
\ell(\mathbf{I}^*, \mathbf{R}^* | \boldsymbol{\Theta}) &= \sum_{t = 0}^{\tau}  
    \Bigg[\log \binom{S_t}{I^*_t} + I^*_t \log \pi_t^{(SI)} + \left(S_t-I^*_t \right) \log \left(1-\pi_t^{(SI)} \right)\\
&\hspace{9mm} + \log \binom{I_t}{R^*_t} + R^*_t \log \pi^{(IR)} + \left(I_t-R^*_t \right) \log \left(1-\pi^{(IR)} \right) \Bigg],
\end{split}
\end{equation}
where the parameter vector $\boldsymbol{\Theta}$ contains $\beta$ and $\gamma$ for the traditional model, and includes additional parameters used to estimated $\beta_t$ for the time-varying transmission models and the BC model. Complete data would provide the time series over the course of the epidemic for the transition vectors $\mathbf{I}^*$ and $\mathbf{R}^*$, the initial conditions $S_0$ and $I_0$, and the population size $N$. Often, we do not observe complete information on infectious and removal times. Unless otherwise stated, we assume that incidence ($\mathbf{I}^*$) is observed, and removals ($\mathbf{R}^*$) must be imputed, using data-augmented MCMC methods \citep{o1999bayesian, lekone}. The R package \texttt{nimble} \citep{nimble-article:2017, nimble-software:2021} was used for computation, as it offers a mechanism for implementing user-defined data-augmented MCMC algorithms.

In the Bayesian framework, the parameter vector must be assigned a prior, with the use and justification of informative priors for any parameters varying by disease application. Often, the gamma distribution is used to specify the prior for $\beta$ and $\gamma$ as both parameters must be positive. In the presence of knowledge about the duration of the infectious period, informative priors can be used for $\gamma$. Determining informative priors for the parameters of the alarm functions is challenging. We use a vague gamma prior for $k$ in the power alarm and Uniform$(0, 1)$ priors for $\delta$ in the threshold and Hill alarms. For $H$ and $x_0$ Uniform$(\min(x), \max(x))$ priors are used as we expect the change point or half occupation point to occur during the observed range of incidence. The spline coefficients $\boldsymbol{b}$ are given vague $N(0, 100)$ priors and the knots are given Uniform$(\min(x), \max(x))$ priors. The parameters of the covariance in the Gaussian process model, $\sigma$ and $l$ are weakly identified \citep{zhang2004inconsistent}, making the use of informative priors crucial. As the alarm function is estimated on the logit scale, the variability of the function is limited, so we use a gamma$(150, 50)$ prior for $\sigma$. The prior for the length-scale parameter was specified as inverse gamma with the shape and scale parameters determined using the practical range approach \citep{gelfand2005bayesian}. Using this approach, the mean of the prior for $l$ is specified by finding the value such that the covariance function is 0.05 for two points that are separated by half the maximum distance observed in the data. The prior standard deviation was fixed at two, as that was found to produce reasonable estimation.

\section{Simulation Study}
\label{simulation}

\subsection{Simulation set-up}

The statistical properties of the BC model are assessed via simulation. The primary goal of the simulation study was determining whether the behavioral change mechanism could be recovered through estimation of the alarm function. The secondary objective was comparing the BC model to the traditional approach without behavioral change and a flexible time-varying transmission model, assessing posterior predictive forecasting and model fit. These aims were addressed by simulating epidemics with behavioral change under the three alarm functions described in Section \ref{methods:bcmodel}. For each of the three data generation scenarios, 50 epidemics were simulated using the initial conditions $N = 1,000,000$, $S_0 = 999,995$, and $I_0 = 5$. Five models were fit to each simulated epidemic: a BC model using the true alarm function, BC models using the spline and Gaussian process alarms, the model with no behavioral change, and a time-varying transmission model with $\beta_t$ estimated flexibly using natural cubic splines. 

To evaluate posterior prediction, epidemics were simulated over 100 days, with the first 50 days used for model fitting  and the subsequent 50 days used to evaluate forecasting accuracy. Simulation parameters were chosen to produce epidemics with a distinguished peak during the first 50 days of the epidemic and with additional peak(s) occurring in the subsequent 50 days. Complete specification of simulation parameters is provided in Supplementary Table 1. Epidemics were generated and BC models were fitted using both 14-day and 30-day average incidence to inform the alarm function. Similar conclusions were found in both settings, so we detail the 30-day average results here and provide the 14-day average results in the Supplementary Material. 

Priors were specified as described in Section \ref{methods:estimation}. For each model, three MCMC chains were run using various starting values of the parameters. The BC models and the flexible $\beta_t$ model required more burn-in iterations due to their increased complexity. All models were run for 300,000 iterations post burn-in with samples drawn every 10th iteration. Full descriptions of priors used and MCMC specifications are provided in Supplementary Table 2. Convergence was established by a Gelman and Rubin diagnostic value below 1.1 \citep{gelmanrubin1992}. A small number of models did not converge despite running for a large number of iterations and have been excluded from the results. More information on these models is provided in the Supplementary Material.

\subsection{Simulation results}

\subsubsection{Alarm function estimation}

\begin{figure}[ht!]
\centering\includegraphics[width = 0.9\textwidth]{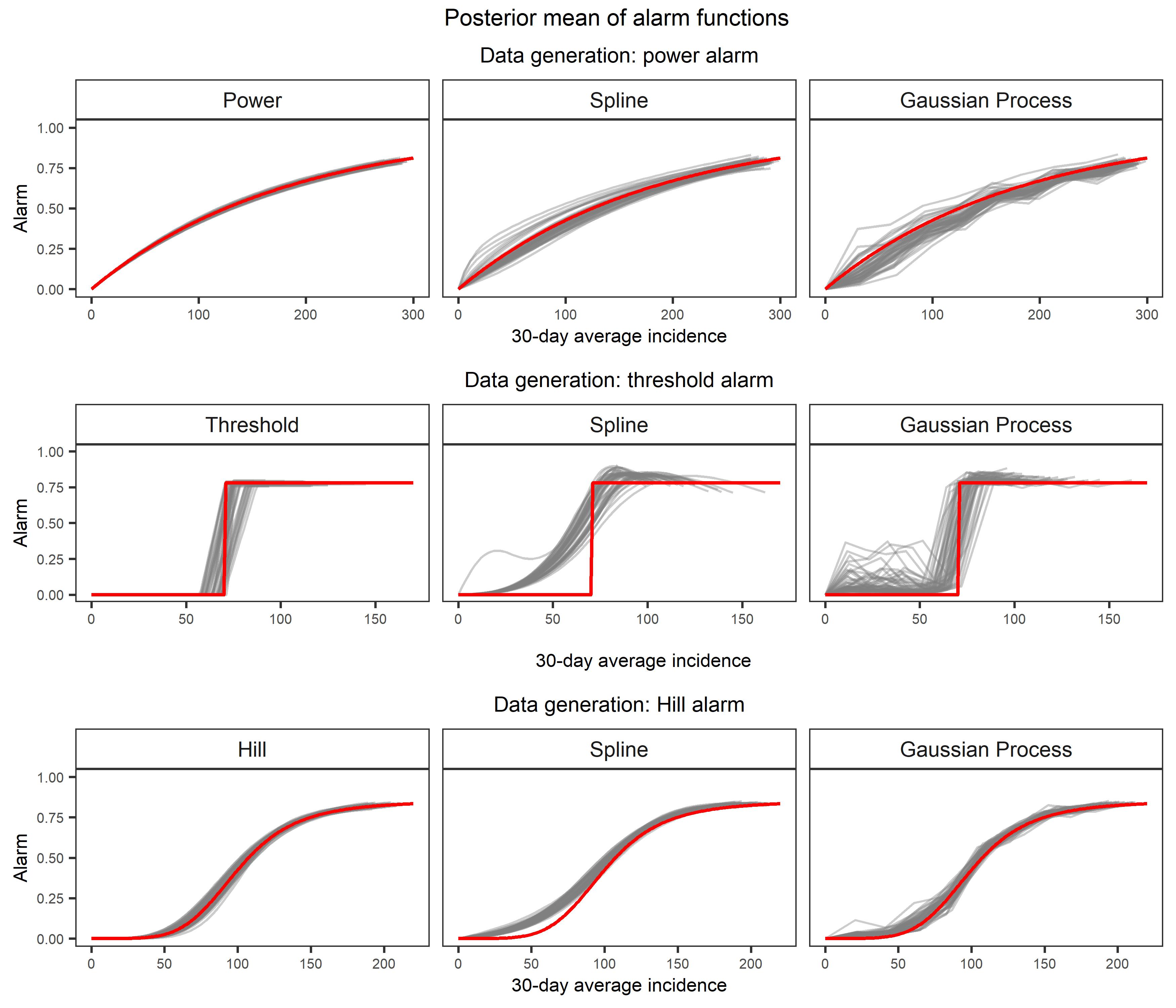}
\caption{True and posterior mean estimates of alarm functions from 50 simulated epidemics using the correct parametric alarm function, and the spline and Gaussian process alarms for model fitting. True alarm functions shown in red.}
\label{Fig3}
\end{figure}

To assess estimation of the alarm function in each data generating scenario, we compare posterior mean alarm function estimates to the true alarm functions (Figure \ref{Fig3}). The alarm is shown as a function of the 30-day average incidence, ranging from zero to the maximum observed value during the epidemic, which varies between simulations. We find estimation of the alarm function to be excellent when the true functional form of the alarm was used in model fitting. More interestingly, we find the spline and Gaussian process approaches recover the alarm function reasonably well, particularly for the power and Hill alarms. The alarm function recovery is not as precise when the threshold alarm is the true function. However, this is expected as the piecewise constant form of this alarm is generally not well described by splines or Gaussian processes, which are inherently smooth. Despite this, both approaches are able to detect the alarm function rapidly increasing and leveling off quite impressively. These results indicate that when analyzing real data where the true alarm function is unknown, the spline and Gaussian process alarms offer robust and flexible possibilities.

\subsubsection{Posterior prediction}

\begin{figure}[ht!]
\centering\includegraphics[width = 0.8\textwidth]{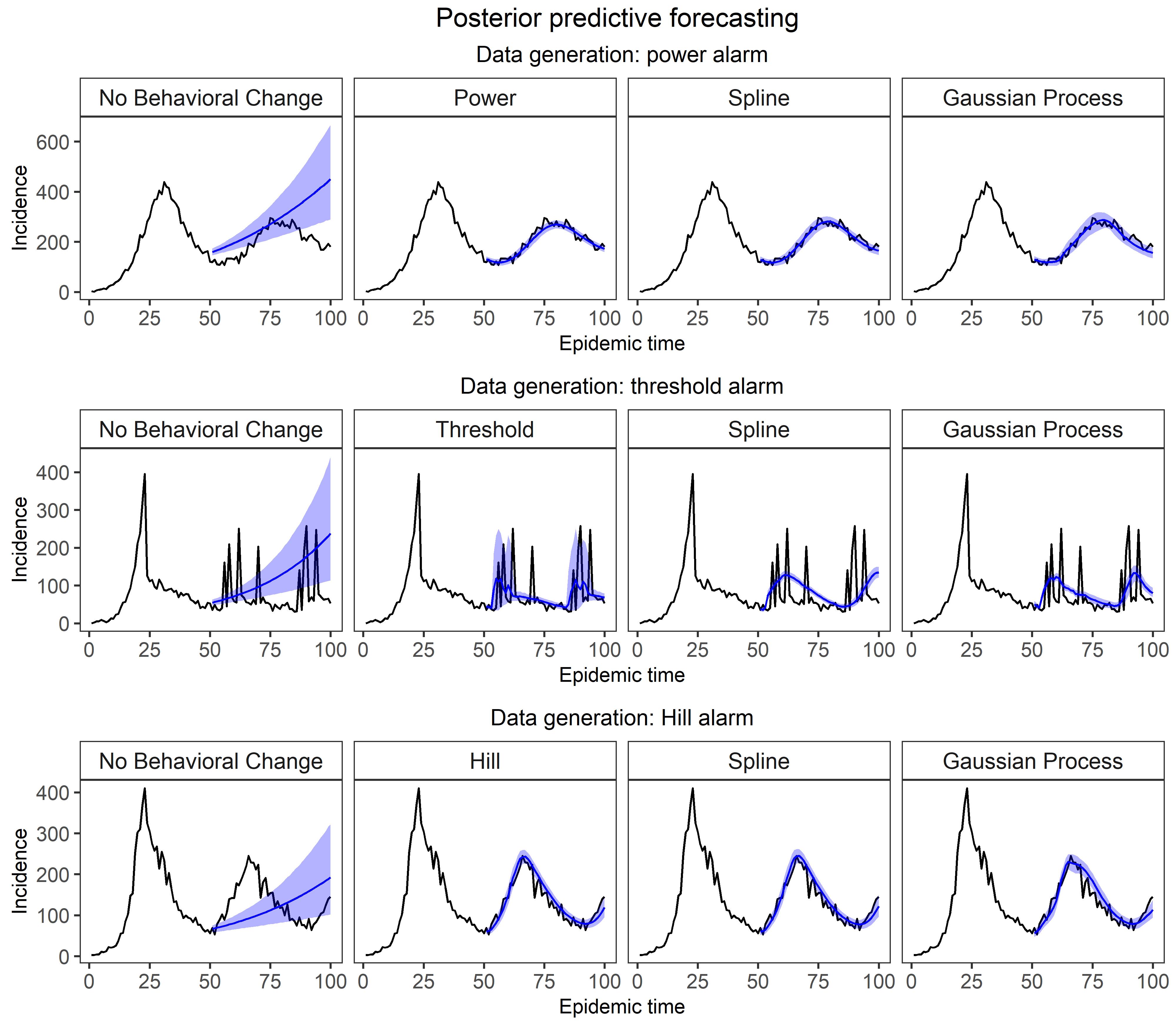}
\caption{Mean and 95\% credible intervals for posterior predictive forecasts of future incidence compared to the truth for a randomly selected simulation.}
\label{Fig4}
\end{figure}

The ability of the fitted BC models to predict the epidemic curve was carried out using the posterior distribution for model parameters derived using the first 50 days of the epidemic. For 10,000 posterior draws of the parameters, the future epidemic trajectory was simulated using the model state on day 50 to determine the initial values and proceeding with binomial draws from $S_t$ and $I_t$ for $t = 51, ..., 100$. A drawback of the flexible $\beta_t$ model is there is no obvious mechanism for forecasting, so we compare the model with no behavioral change to the three BC models (true, spline, and Gaussian process alarms). Consistent results were found across simulations. For brevity, the posterior predictive distribution is provided for a single, randomly selected, and typical simulation in Figure \ref{Fig4}.

The model which does not incorporate behavioral change does a poor job of posterior prediction. Assuming constant transmission yields an estimated $\beta$ which averages over what was observed during the first 50 days. Correspondingly, the posterior predictive trajectories are increasing, but at a slower rate than the growth observed at the start of the epidemic. In contrast, posterior predictions from the BC models are able to detect the intensity and timing of subsequent waves of transmission. Minimal differences were found between predictions from the parametric and non-parametric approaches for data generated with the power and Hill alarms. For epidemics simulated using the threshold alarm, the non-parametric approaches predict smoother subsequent peaks, due to their failure to capture the abruptness of the change point in the alarm. Finally, the 95\% posterior credible intervals for the model with no behavioral change widen as predictions become further from the last observed time point. This is not seen for the BC models, as the structural dependency encoded by the alarm function restricts the shape of the epidemic curve.

\subsubsection{Comparing model fit}

Model fit was assessed with the Widely Applicable Information Criteria (WAIC) \citep{watanabe2010asymptotic}, for which lower values indicate superior model fit. The distribution of WAIC values and the proportion of epidemics where each model had the lowest WAIC are summarized in Table \ref{Table1}. The true alarm function was selected 82 - 96\% of the time, indicating that WAIC is able to correctly identify the best model. The spline and Gaussian process BC models sometimes provided the best fit. This was less likely when the data was generated from the threshold alarm, where recovery of the alarm function by these methods was poorest. Notably, despite its flexibility, the time-varying $\beta_t$ model was only selected once and generally had higher average WAIC than any of the BC models. The exception was for data generated from the power alarm, the most gradual alarm function. This indicates that the additional structure imposed by the BC models results in lower WAIC in the simulation setting where data was generated from a model with behavioral change present. The traditional model with no behavioral change performs poorly across all three data generation scenarios.

\begin{table}[ht!]
\small\sf\centering
\caption{Summaries of WAIC values across 50 simulated epidemics from three data generation scenarios. Models are ordered by mean WAIC.  \label{Table1}}
\centering
{\tabcolsep=4.25pt
\begin{tabular}{llcc}
\toprule
\textbf{Data generation} & \textbf{Model fitted} & \makecell[c]{\textbf{WAIC}\\\textbf{Mean (SD)}} & \textbf{\% selected}\\
\midrule
 & Power & 369.63 (12.80) & 92\%\\

 & Spline & 371.43 (12.73) & 8\%\\

 & $\beta_t$ & 375.28 (13.19) & 0\%\\

 & Gaussian process & 381.44 (13.78) & 0\%\\

\multirow{-5}{*}{\raggedright\arraybackslash Power} & No Behavioral Change & 673.74 (16.25) & 0\%\\
\cmidrule{1-4}
 & Threshold & 333.19 (12.61) & 96\%\\

 & Gaussian process & 367.06 (48.96) & 4\%\\

 & Spline & 484.39 (65.67) & 0\%\\

 & $\beta_t$ & 720.61 (59.87) & 0\%\\

\multirow{-5}{*}{\raggedright\arraybackslash Threshold} & No Behavioral Change & 1058.45 (120.73) & 0\%\\
\cmidrule{1-4}
 & Hill & 358.55 (13.61) & 82\%\\

 & Spline & 360.28 (13.33) & 10\%\\

 & Gaussian process & 364.06 (13.95) & 6\%\\

 & $\beta_t$ & 375.39 (19.04) & 2\%\\

\multirow{-5}{*}{\raggedright\arraybackslash Hill} & No Behavioral Change & 793.40 (34.11) & 0\%\\
\bottomrule
\end{tabular}}
\end{table}   

\section{Data Applications}\label{dataAnalysis}

\subsection{Ebola Disease}

\begin{figure}[ht!]
\centering
\includegraphics[width = 0.8\textwidth]{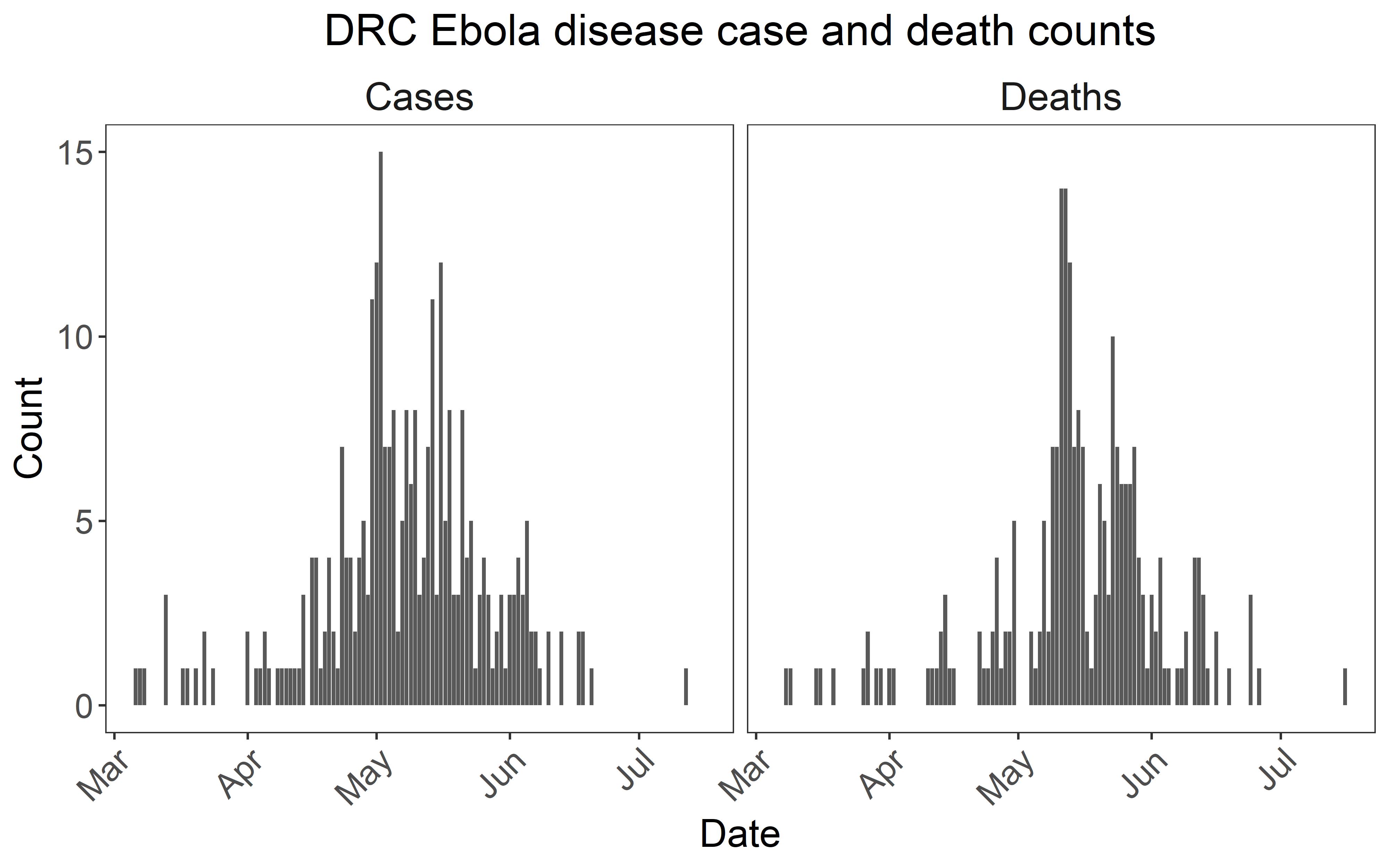}
\caption{Democratic Republic of the Congo observed Ebola disease data. Cases are recorded by symptom onset date and deaths by death date.}
\label{Fig5}
\end{figure}

We first illustrate the BC model on a well-studied Ebola outbreak which occurred in 1995 in the Democratic Republic of the Congo (DRC). Ebola is a deadly disease which transmits between humans through direct physical contact with bodily fluids or contaminated clothes or bedding \cite{cdcEbola}. A person is only infectious once they develop signs and symptoms of Ebola disease, which can occur anywhere between two to 21 days (average eight to 10 days) after initial contact with an ebolavirus \cite{cdcEbola, cdcEbolaSymptoms}. Once symptoms have appeared, individuals remain infectious for four to ten days and may either recover or die, with the average case fatality rate around 50\% \cite{whoEbola}. The 1995 DRC epidemic occurred primarily in the city of Kikwit in the Bandundu region, which had a population of 5,363,500 during the outbreak \cite{lekone}. The data used in this analysis are publicly available in the \texttt{outbreaks} R
package \cite{outbreaks} and contain symptom onset date for 291 cases and death date for 236 individuals documented between March and July 1995 (Figure \ref{Fig5}). It is known that 316 infections occurred, but symptom onset date was not recorded for 25 individuals. Ebolavirus was identified as the causative agent of the outbreak on May 9th, after which control measures were immediately introduced. Further details about this epidemic can be found in Khan et al. (1999) \cite{khan1999reemergence}.

Due to the long latent period for Ebola disease, we extend the SIR model introduced in Section \ref{methods} to incorporate an additional Exposed compartment describing individuals in the period between contracting the Ebolavirus and having symptoms. The SEIR model is a direct extension to the SIR model for which we now define $E_t$ as the number of individuals in the exposed compartment and $E^*_{t}$ as the number of newly exposed individuals during the continuous time interval $[t, t+1)$. The difference equations become:
\begin{eqnarray*} 
    S_{t+1} &=& S_t - E^*_t  \\ 
    E_{t+1} &=& E_t + E^*_t - I^*_t  \\ 
    I_{t+1} &=& I_t + I^*_t - R^*_t  \\ 
    R_{t+1} &=& R_t + R^*_t.
\end{eqnarray*}
The transitions between compartments are still assumed to be binomially distributed as $E^*_{t} \sim Bin\left(S_{t}, \pi_t^{(SE)} \right)$, $I^*_{t} \sim Bin\left(E_{t}, \pi^{(EI)} \right)$ and $R^*_{t} \sim Bin\left(I_{t}, \pi^{(IR)} \right)$. Now, $\pi_t^{(SE)}$ is the transmission probability of interest, as it describes the probability of an infectious individual transmitting the pathogen to a susceptible individual. As before, the form of the transmission probability can be described by the right hand side of Equations \ref{eq:popAvgtransprob}, \ref{eq:pSIBetaT}, or \ref{eq:BCMtransprob}, depending on whether time-varying transmission is allowed or behavioral change is captured with an alarm function. The other transition probabilities are specified as $\pi^{(EI)} = 1 - \exp(-\lambda)$ and  $\pi^{(IR)} = 1 - \exp(-\gamma)$, where $\lambda$ and $\gamma$ are the mean lengths of the latent and infectious periods, respectively. Here, $\boldsymbol{I}^*$ and $\boldsymbol{R}^*$ are partially defined by the data, and $\boldsymbol{E}^*$ is completely unobserved. Missing exposure, symptom onset, and removal dates are imputed using data-augmented MCMC methods \citep{lekone}.

Using the SEIR framework, various models were fitted. Five BC models using the power, threshold, Hill, spline, and Gaussian process alarms were evaluated. For comparison, we also fit models with no behavioral change, a model with a pre-specified intervention effect, and the flexible $\beta_t$ model as described in the simulation study. Previous modeling of this epidemic has evaluated a pre-specified intervention effect by assuming constant transmission until the time of the intervention, after which transmission decays exponentially \cite{lekone, ward2023idd}. Mathematically, this is written by allowing $\beta_t$ in Equation \ref{eq:pSIBetaT} to be equal to $\exp[\beta_1 + \beta_2 (t - t^*) \mathbbm{1}(t \geq t^*)]$, where $\beta_1$ corresponds to the baseline intensity, and $\beta_2$ represents the decay in transmission after the intervention was introduced at time $t^*$ (May 9th).

In the BC models, we let the alarm be informed by cumulative observed incidence as this epidemic was completely observed. Although there were 25 additional infections, as the timings of these are unknown, so we do not allow them to influence the population alarms. The first three observed cases occurred consecutively on March 6, 7, 8, so we start our modeling on March 8th and assume $E_0 = 2$ and $I_0 = 1$. It's known that two deaths occurred prior to this date \cite{khan1999reemergence}, so we set $R_0 = 2$. We assume $N = 5,363,500$ \cite{lekone}. Vague priors for $\beta$ and the parameters of the alarm function and flexible spline model were specified as in Section \ref{methods:estimation}. The priors for $\lambda$ and $\gamma$ were informative and specified as in previous modeling of this epidemic \cite{lekone}. Full specification of all priors for model parameters are detailed in Supplementary Table 12. Convergence was established by a Gelman and Rubin diagnostic value below 1.1 \citep{gelmanrubin1992}. 

WAIC was used to determine the best fitting models (Table \ref{Table2}). All five BC models had lower WAIC than the standard models used for comparison, indicating the BC models have better fit than the other approaches. Interestingly, the parametric threshold and Hill BC models had lower WAIC than the non-parametric Gaussian process and spline models, with the threshold model having the overall lowest WAIC. Of the standard approaches, WAIC was relatively similar between the flexible $\beta_t$ model and the model with pre-specified change in transmission after the intervention. The no behavioral change model has the highest WAIC indicating poor fit.

\begin{table}[ht!]
\small\sf\centering
\caption{WAIC values for all models used in the DRC Ebola disease analysis. Models are ordered by WAIC. \label{Table2}}
    \begin{tabular}{lcc}
        \toprule
        \textbf{Type} & \textbf{Model fitted} & \textbf{WAIC}\\
        \midrule
         & Threshold & 600,676.6\\
        
         & Hill & 603,614.8\\
        
         & Gaussian Process & 605,838.9\\
        
         & Power & 611,096.6\\
        
        \multirow{-5}{*}{\raggedright\arraybackslash BC Model} & Spline & 616,479.3\\
        \cmidrule{1-3}
         & $\beta_t$ & 622,297.5\\
        
         & Intervention & 622,767.5\\
        
        \multirow{-3}{*}{\raggedright\arraybackslash Standard Approach} & No Behavior Change & 675,743.6\\
        \bottomrule
    \end{tabular}
\end{table}

\begin{figure}[ht!]
\centering
\includegraphics[width = \textwidth]{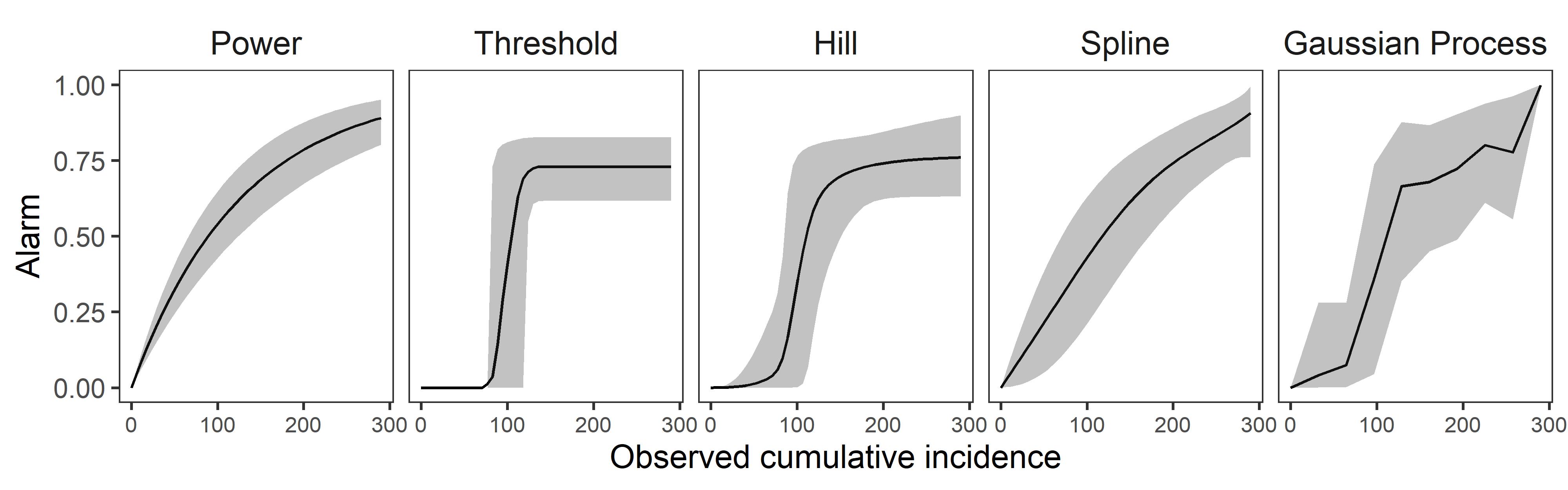}
\caption{Posterior means and 95\% credible intervals for all estimated alarm functions for the DRC Ebola outbreak.}
\label{Fig6}
\end{figure}

Figure \ref{Fig6} depicts the estimated alarm functions for all the BC models. The power, spline, and Gaussian process alarms are relatively similar in shape, with steady increase in alarm as more cases were observed during the epidemic, with peak alarm around 0.8-1.0. On the other hand, the threshold and Hill alarms both estimate low alarm values until around 100 total cases had been observed, after which the estimated alarm value shifted to around 0.75. The 100th case was observed on May 2nd, which also corresponds to the peak of the epidemic. It's worth noting, however, that this was one week prior to the discovery of Ebola as the cause of the outbreak and subsequent introduction of control measures.

As the threshold BC model provided the best fit per WAIC, we compare estimated reproductive numbers and posterior predictive distributions of cumulative incidence between the BC model and the three standard modeling approaches. The posterior predictive distribution was computed using 10,000 posterior draws of the model parameters, and for each draw, the epidemic trajectory is simulated from the chain binomial model. When fitting the BC models, it was assumed that the alarm was a function of only the 291 observed cases, as the symptom onset date was missing for the other 25. To ensure accuracy in the posterior predictive distribution for the BC models, we assumed only 92\% of cases were observed and only observed cases could impact the alarm function.

\begin{figure}[ht!]
\centering
\includegraphics[width = \textwidth]{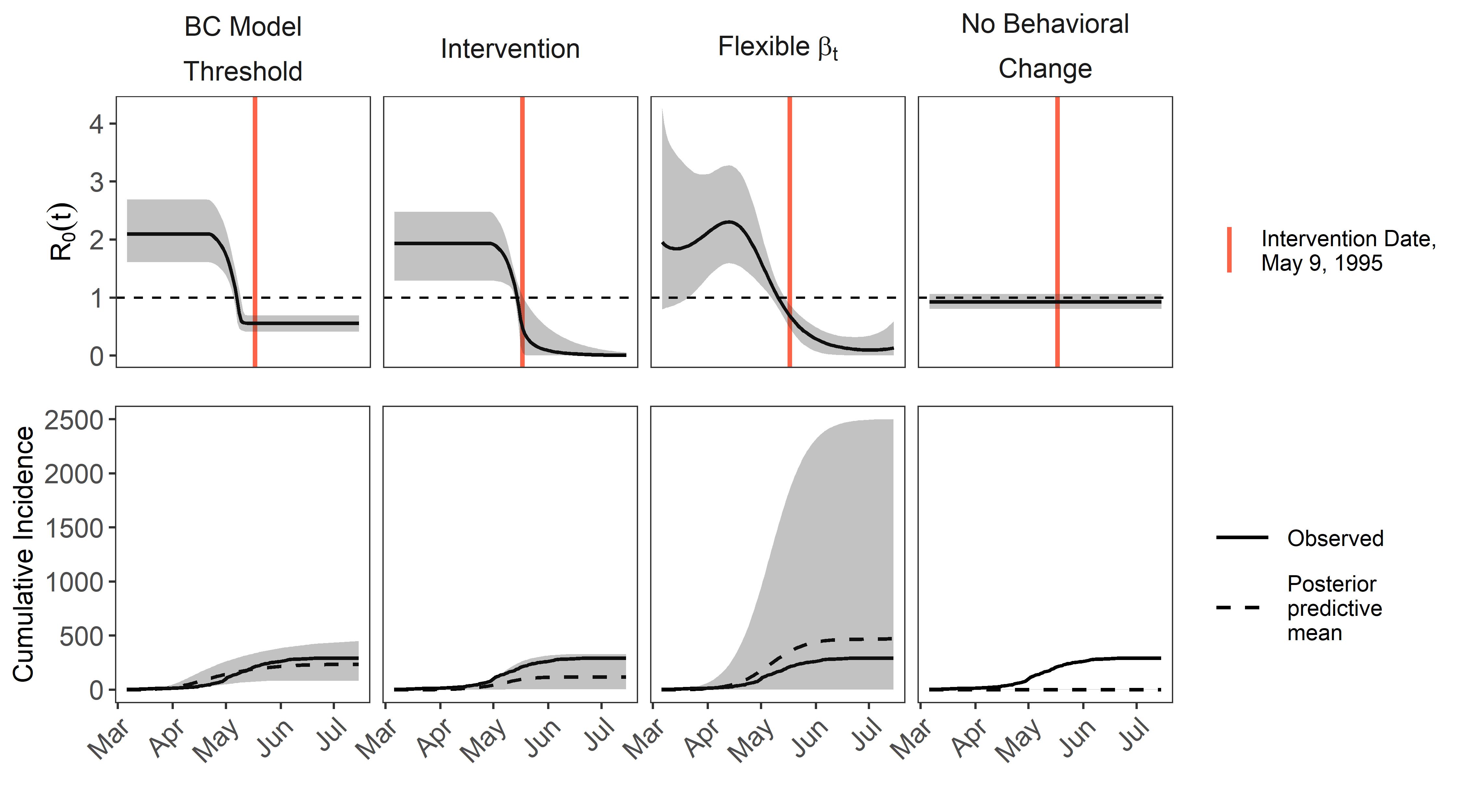}
\caption{Posterior means and 95\% credible intervals for the effective reproductive number over time and the posterior predictive distribution of cumulative incidence for the DRC Ebola outbreak. Results are presented for the BC model using the threshold alarm, the model with an exponential decay intervention, the flexible $\beta_t$ model, and the model with no behavioral change. }
\label{Fig7}
\end{figure}

The BC model, intervention model, and the flexible $\beta_t$ model all provide relatively similar estimates of $\mathcal{R}_0(t)$ at the start of the epidemic with posterior means between 1.9 and 2.1 (Figure \ref{Fig7}). However, the intervention model is restricted in shape as the change point in transmission is fixed at May 9th, and the exponential decay form of the intervention effect forces transmission and $\mathcal{R}_0(t)$ to zero by the end of the epidemic. Conversely, the threshold alarm can detect a change point in transmission which best fits the data, and the alarm level after the change point dictates the reduction in $\mathcal{R}_0(t))$, although $\mathcal{R}_0(t)$ must remain constant after the change point occurs. This flexibility allows the BC model to detect behavioral change resulting in $\mathcal{R}_0(t)$ dropping below 1 on April 30, around a week prior to the implementation of public health interventions. The flexible $\beta_t$ model results in reproductive number estimates similar to the intervention model, although with much wider variance at the start of the epidemic and while the intervention model estimates $\mathcal{R}_0(t) \approx 0$ by the end of May, the flexible $\beta_t$ models estimates $\mathcal{R}_0(t)$ to level out at a value of 0.1. 

The subtle differences in the trajectory of $\mathcal{R}_0(t)$ between the modeling approaches lead to larger differences in the posterior predictive distribution of cumulative incidence. As the reproductive number estimated for the intervention model goes quickly to zero, the final size of the epidemic is underestimated with a posterior predictive mean of 116 compared to the observed value of 316. The flexible $\beta_t$ model overestimates the final size with a posterior predictive mean of 471 and also has huge variability in predicted epidemic trajectory. The threshold BC model has the best posterior predictive distribution although it still underestimates the final epidemic size with the mean at 235. The model with no behavioral change is forced to average transmission over the entire epidemic and estimates $\mathcal{R}_0(t)$ just below 1 over the entire time period, which results in posterior predictive distribution over epidemics that only infect one or two individuals before dying out.

\subsection{COVID-19}

\begin{figure}[ht!]
\centering\includegraphics[width = 0.8\textwidth]{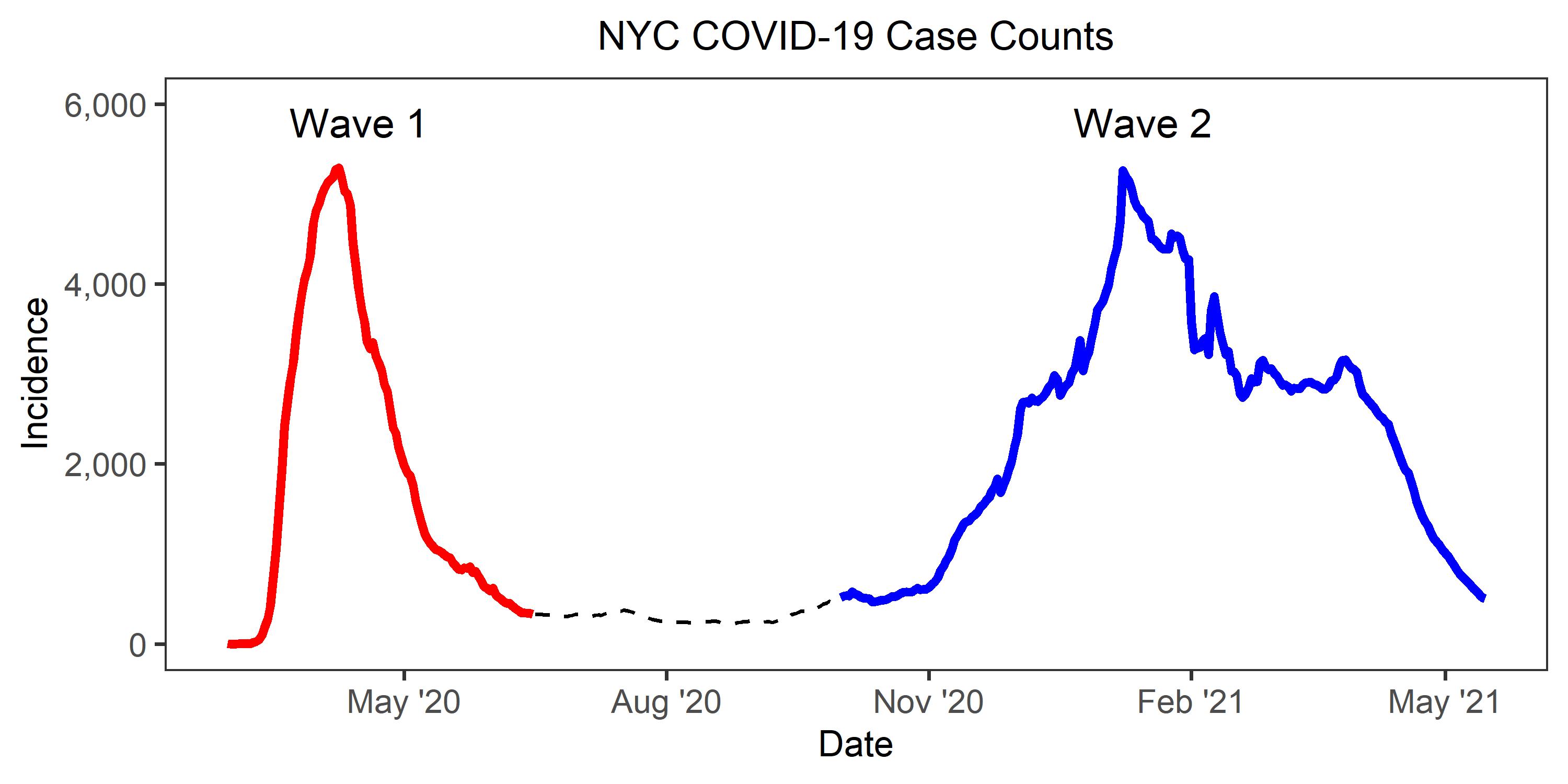}
\caption{New York City COVID-19 data and waves defined for the analysis. Wave 1 occurs between Mar 1 and Jun 15 2020 and Wave 2 between Oct 1, 2020 and May 15, 2021.}
\label{Fig8}
\end{figure}

After the first cases of COVID-19 were identified in China in December 2019, the SARS-CoV-2 virus spread rapidly across the globe, being declared a pandemic by the World Health Organization in March 2020 \citep{whoPandemic}. We illustrate the use of the BC model to evaluate change in behavior over time using two waves of COVID-19 data from New York City (NYC) during the period March 2020 - May 2021. The data used for this study are publicly available as part of the NYC Department of Health and Mental Hygiene Github repository \citep{nycCOVIDgithub}. Based on the observed case counts, we define two waves of COVID-19 in NYC to be analyzed (Figure \ref{Fig8}). 

To illustrate the differences in the various BC model specifications, the power, threshold, Hill, spline, and Gaussian process alarms were all used to model both COVID-19 waves. As in the simulation study, the model with no behavioral change and the flexible $\beta_t$ model were also fitted for comparison purposes. Prior to model fitting, the reported counts of new cases over time were smoothed to the 7-day average to account for weekly fluctuations in reporting. Although COVID-19 is known to have a short latent period around 5 to 6 days \cite{xin2022estimating}, this is difficult to incorporate in a fully Bayesian model due to the lack of reliable data arising from the presence of asymptomatic infections and lag between infectiousness and testing \cite{subramanian2021quantifying}, and therefore the SIR model is commonly used \cite{de2020bayesian, lawson2021space, irons2021estimating, lawson2022bayesian}. For this analysis, we assume the smoothed incidence provides the time-series $\boldsymbol{I^*}$, and use data-augmented MCMC to estimate $\boldsymbol{R^*}$. While we know this is not true because of under-reporting, this assumption reduces computational expense. Additionally, as we anticipate behavioral change to be informed by observed epidemic trajectory, we do not expect the absence of undetected cases in our model to impact the estimation of our alarm function.

The population size was set as $N =$ 8,804,190, the recorded population from the 2020 census. For each wave, we allow the initial conditions $S_0$ and $I_0$ to be estimated, using strong priors based on the past epidemic trajectory. Vague priors for $\beta$ and the parameters of the alarm function and flexible spline model were specified as in Section \ref{methods:estimation}. The prior for $\gamma$ was specified to correspond with a mean infectious period of three days and 80\% prior probability of the mean between 2 and 4 days. This corresponds with the typical length of time between an individual becoming contagious and testing positive and subsequently isolating. Results of a sensitivity analysis on this prior indicated little impact to the main conclusions across various priors and are provided in the Supplementary Material. Full specification of all priors for model parameters are detailed in Supplementary Table 21. In this example, the epidemic was not fully observed, so the BC models were fit using both 30-day and 60-day average incidence to inform the alarm functions. Convergence was established by a Gelman and Rubin diagnostic value below 1.1 \citep{gelmanrubin1992}. 

To determine the best fitting models for the two COVID-19 waves, WAIC values were compared (Table \ref{Table3}). Modeling $\beta_t$ flexibly with splines provided the best fit to both waves, indicating that the additional structure imposed by the alarm function does not fully capture the observed epidemic trajectory. The Gaussian process and Hill alarms were among the BC models with the lowest WAIC across both waves. For the first wave, using 60-day average incidence to inform the alarm function offered better model fit, while the 30-day average performed better for the second wave. This is likely due to the different shape of the epidemic curve during Wave 2, which peaked in the beginning of January 2021, but leveled off between March - April 2021 before incidence was truly driven down. In contrast, Wave 1 showed steady decline in incidence post-peak, indicating the behavioral change in the population continued until the wave died out. The model with no behavioral change had the highest WAIC for Wave 1 and for Wave 2 had the second highest WAIC, indicating the importance of incorporating of behavioral change when considering real epidemic data. 

\begin{table}[ht!]
\small\sf\centering
\caption{WAIC values for all converged models used in the NYC COVID-19 analysis. Within each wave, models are ordered by WAIC. The BC models chosen for the final results are indicated in italics. The power alarm models did not converge for Wave 2 and are therefore excluded from these results.  \label{Table3}}
\centering
{\tabcolsep=4.25pt
    \begin{tabular}{lccc}
\toprule
\textbf{Wave} & \textbf{Model fitted} & \textbf{Smoothing} & \textbf{WAIC}\\
\midrule
 & $\beta_t$ & None & 1089.83\\

 & Gaussian Process & 60-day & 1092.92\\

 & Spline & 60-day & 1099.02\\

 & Hill & 60-day & 1125.36\\

 & Spline & 30-day & 1157.33\\

 & Gaussian Process & 30-day & 1157.89\\

 & Hill & 30-day & 1190.37\\

 & Threshold & 30-day & 1466.06\\

 & Threshold & 60-day & 1466.34\\

 & Power & 60-day & 1540.83\\

 & Power & 30-day & 1753.43\\

\multirow{-12}{*}{\raggedright\arraybackslash Wave 1} & No Behavior Change & None & 2263.96\\
\cmidrule{1-4}
 & $\beta_t$ & None & 2865.70\\

 & Gaussian Process & 30-day & 3025.92\\

 & Threshold & 30-day & 3041.86\\

 & Hill & 30-day & 3045.90\\

 & Spline & 30-day & 3051.68\\

 & Threshold & 60-day & 3078.72\\

 & Hill & 60-day & 3080.96\\

 & Gaussian Process & 60-day & 3091.50\\

 & No Behavior Change & None & 3102.84\\

\multirow{-12}{*}{\raggedright\arraybackslash Wave 2} & Spline & 60-day & 3105.28\\
\bottomrule
\end{tabular}}
\end{table}

\begin{figure}[ht!]
\centering\includegraphics[width = 0.9\textwidth]{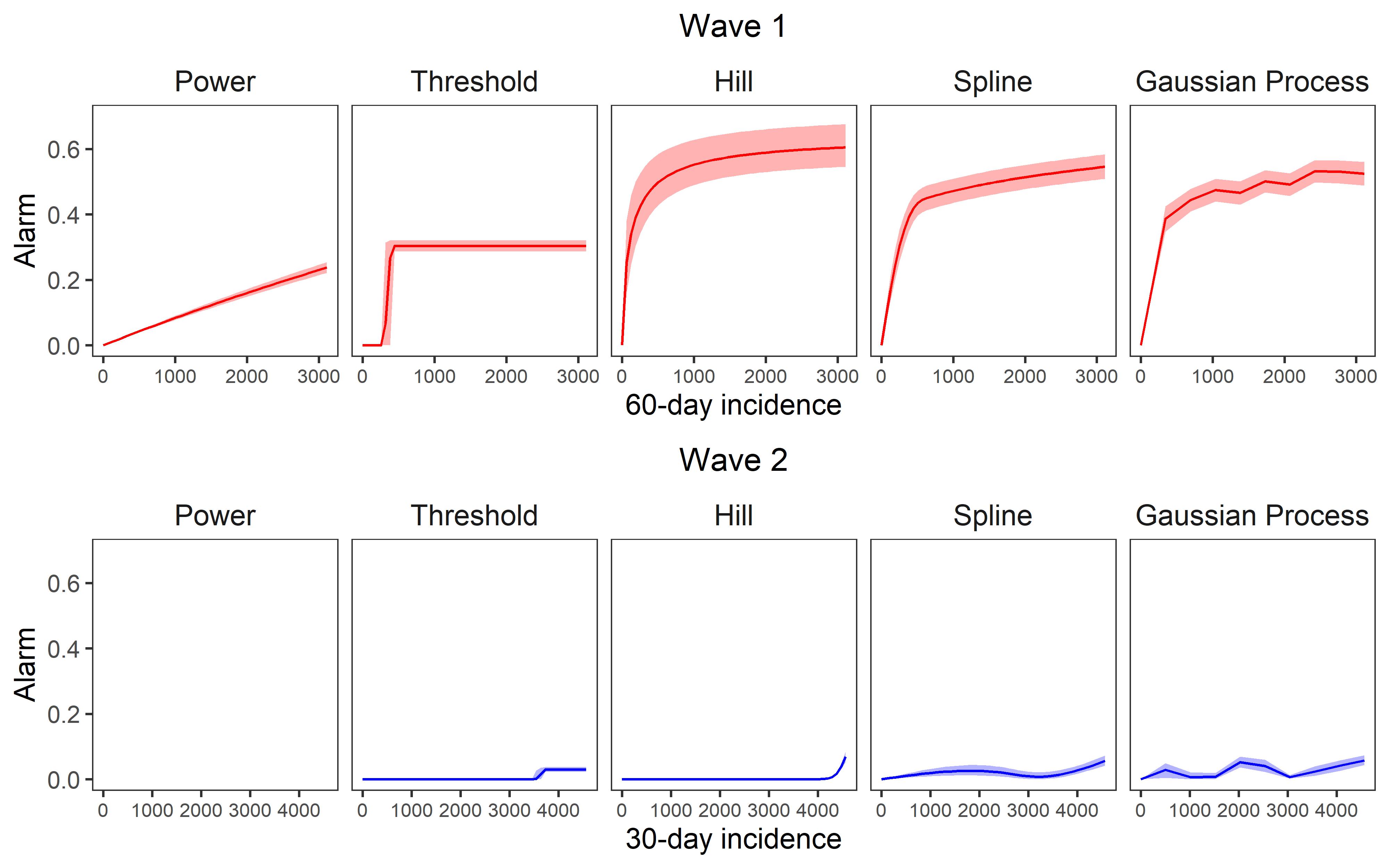}
\caption{Posterior means and 95\% credible intervals for all estimated alarm functions from each wave of the NYC COVID-19 epidemic. The power alarm model did not converge for Wave 2 and is therefore excluded from these results. }
\label{Fig9}
\end{figure}

Based on the WAIC results, we present the estimated alarm functions for each BC model based on 60-day average incidence for Wave 1 and 30-day average incidence for Wave 2 (Figure \ref{Fig9}). Despite the restricted shapes of the parametric functions, the estimated alarm functions were generally similar. The spline and Gaussian process alarms were very alike, which is not surprising due to the relationship between Gaussian processes and splines \citep{wahba1978improper}. Comparing the estimated alarm functions between the two waves can be used to evaluate changes in pandemic response over time. During the first wave, the alarm reached very high levels at relatively low observed incidence. The second wave started while many restrictions from the first wave were still in place, and correspondingly the BC models estimate very slight increases in alarm at higher levels of observed incidence. In the second wave, the estimated Gaussian process alarm is not monotonically increasing which allows it to capture raised alarm during the large winter peak of the epidemic and during the subsequent mini-peak occurring in the spring. As the Gaussian process alarm achieved the lowest WAIC, it seems this flexibility allows for better model fit.

\begin{figure}[ht!]
\centering\includegraphics[width = 0.95\textwidth]{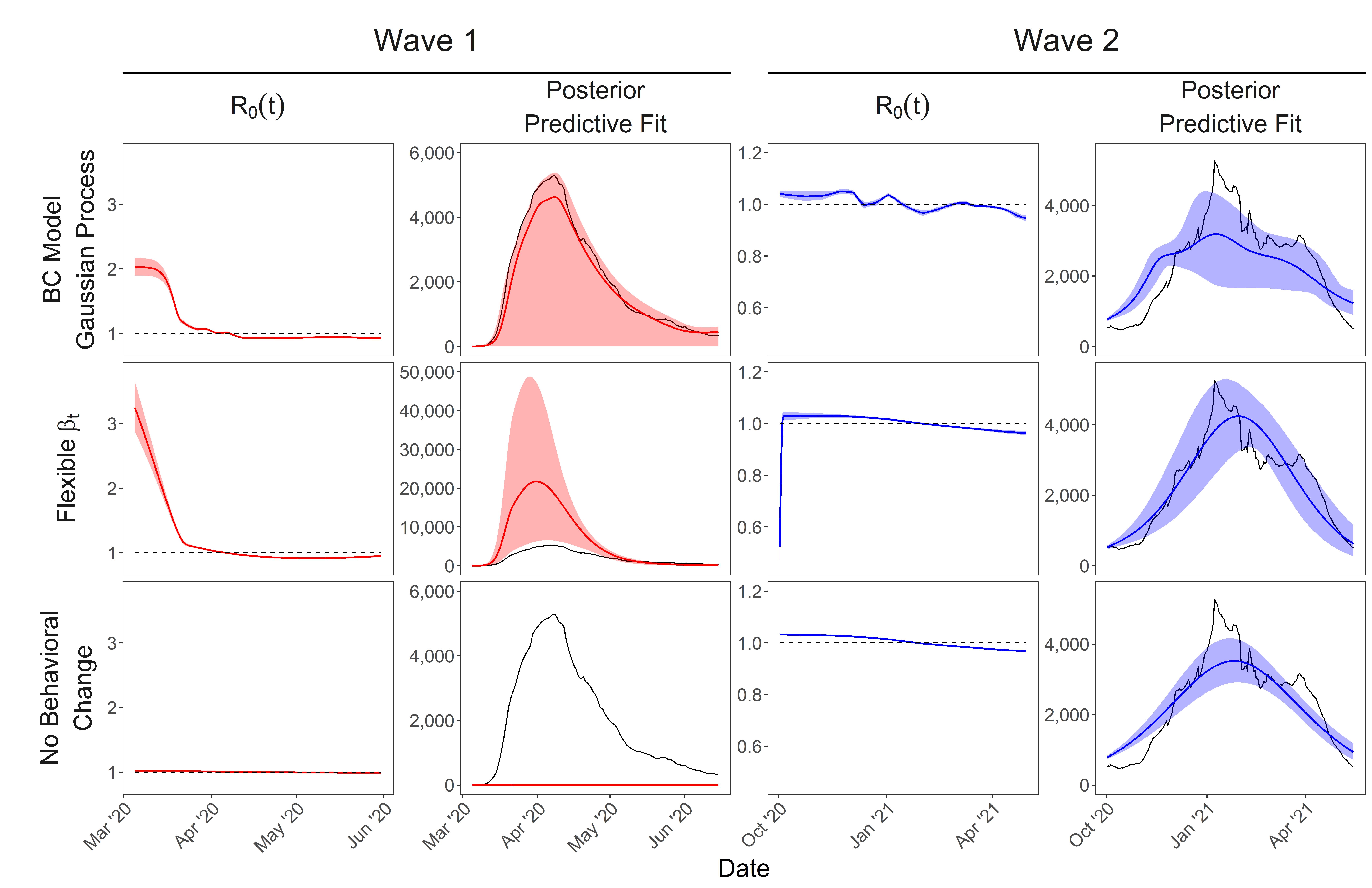}
\caption{Posterior means and 95\% credible intervals for the effective reproductive number of time and the posterior predictive distribution from each wave of the NYC COVID-19 epidemic. Results are presented for the BC model using the Gaussian process alarm, the flexible $\beta_t$ model and the model with no behavioral change. }
\label{Fig10}
\end{figure}

Finally, we compare estimated reproductive numbers and posterior predictive distributions between the BC model using the Gaussian process alarm, the flexible $\beta_t$ model, and the model with no behavioral change (Figure \ref{Fig10}). The Gaussian process alarm was chosen as it had the lowest WAIC of the BC models for both waves. The posterior predictive distribution was computed using 10,000 posterior draws of the model parameters, which includes $S_0$ and $I_0$. For each draw, the epidemic trajectory is simulated from the chain binomial model and is compared to the true observed epidemic curve. For the first COVID-19 wave, the trajectory of $\mathcal{R}_0(t)$ is quite different between the three models, particularly between March and April 2020. The BC model estimates $\mathcal{R}_0(t)$ starting around 2, while the flexible $\beta_t$ model estimates a higher value around 3, and the model with no behavioral change estimating $\mathcal{R}_0(t) \approx 1$ over the entire wave. This leads to vastly different posterior predictive distributions, due to the high variability in stochastic epidemic models at the start of an epidemic. Interestingly, the flexible $\beta_t$ model has poor posterior predictive fit, despite having the lowest WAIC. This is likely because WAIC weights each time point equally in calculating the log predictive density, whereas the posterior predictive epidemic trajectory is highly influenced by the estimated $\mathcal{R}_0(t)$ at time one. In Wave 2, the estimated $\mathcal{R}_0(t)$ for the BC model and the no behavioral change model is just above 1 until mid-January 2021. The flexible $\beta_t$ model follows a similar trajectory, except $\mathcal{R}_0(t)$ is below 1 for the first two days, which allows it to better capture the slow growth of the epidemic at the beginning of October 2020 in the posterior predictive trajectory. The BC model struggles with the shape of the epidemic curve in Wave 2, as incidence stops declining between March and April 2021. However, the posterior predictive distribution appears slightly better than that of the model with no behavioral change.

\section{Discussion}\label{discussion}

There is a critical need to understand the dynamics of population behavior changing in response to an infectious disease outbreak. Guided by the previous deterministic literature, we developed a novel Bayesian epidemic model framework which characterizes behavioral change dynamics at the population level while remaining simple enough to be computationally feasible. We showed that the proposed BC model can accurately estimate the mechanism of behavioral change across a wide range of scenarios, including when flexible non-parametric methods are used. The practical implications and usefulness of the proposed approach were illustrated with two relevant case studies using Ebola and COVID-19 data, although the model could be applied to any communicable disease.

Our simulation study conducted a thorough investigation of the BC model and made several notable findings. First, when behavioral change impacts epidemic trajectory, the BC model is able to accurately estimate the mechanism of behavioral change. We considered three different alarm functions to describe behavioral change, and although the functions impact the epidemic trajectory differently, we were able to reasonably capture all three functions using non-parametric splines and Gaussian processes. This is hugely beneficial, as it may be difficult to choose an appropriate functional form of the alarm when analyzing real epidemic data. We also showed that posterior predictive forecasts from the BC model can accurately detect a second peak, something that is not feasible with the traditional SIR model. Finally, WAIC provided an accurate metric for selecting the best fit model and we found the additional structure in the BC models generally lead to lower WAIC values than the more flexible approach of estimating $\beta_t$ directly, even though both approaches are able to capture changes in transmission over time.

Our analysis of the Ebola and COVID-19 epidemics offers numerous insights into behavioral change. In both analyses, we found the incorporation of behavioral change offers superior model fit compared to an approach without behavioral change. In the analysis of the Ebola outbreak, we illustrated the ability of the BC model to detect behavioral change which occurs separately from a government intervention. Allowing for structured behavioral change without restricting the timing or impact offered superior model fit and provided additional insight on the population engaging in protective behaviors a week prior to Ebola being identified as the cause of the outbreak. The use of the SEIR model in this analysis also shows how the BC model can be incorporated into more complex compartmental models when additional data is available. In the analysis of the first two waves of the COVID-19 pandemic in New York City, we presented the use of the BC model for comparing behavioral change over time.

Although illustrating the proposed model on real epidemic is extremely valuable, these analyses are not without limitations. In particular, the COVID-19 pandemic has led to many modeling challenges arising from the large presence of asymptomatic cases, lack of testing availability, changes in disease severity over time, and waning immunity. By modeling the COVID-19 waves separately, we mitigate issues with changing disease severity and waning immunity. We have not incorporated undetected infections in the model, however, some preliminary simulations (not shown) have found that the presence of undetected infections which are not accounted for in the model does not impact the estimation of the alarm function when the alarm is based on observed cases, a realistic assumption. While this model misspecification does likely lead to underestimation of the reproductive number, our primary goal in this work is to illustrating the BC models ability to estimate behavioral change and we believe this is accomplished with the presented model. Various methods to account for undetected infections exist, including the addition of vaccinated or asymptomatic compartments \cite{angeli2022modeling} or incorporating sero-prevalence surveys \cite{irons2021estimating}. The BC model could be directly incorporated into these more complex structures when needed to achieve the analysis goals. 

In this work we considered a population-average model which assumes homogeneous mixing and equal susceptibility for all members of the population. These assumptions may not be realistic, as factors like age are known to impact contact rates and susceptibility. Many extensions to the Bayesian SIR model have been introduced to relax these assumptions, including the use of a stratified population structure \citep{porter2016spatial, brownReproductive}, the addition of spatial random effects in the transmission rate \citep{lawson2021space, mahsin2022geographically}, or allowing the transmission rate to be impacted by individual-level covariates \citep{deardon2010}. Establishing the BC model in the population-averaged framework is important as data are often limited, but ongoing work seeks to incorporate behavioral change into models with individual covariates and spatial structure. In a more precise model where susceptibility and contact patterns are estimated separately, the alarm function could modify either factor. The alarm function itself may also be modeled as a function of individual-level covariates. Additionally, one may consider both the transmission rate and the alarm function to vary spatially, and models incorporating spatial structure in the alarm function could estimate region-specific alarms.

Across all alarm functions considered, we assume the alarm level is zero when there have been no cases observed. This assumption is realistic for new or localized outbreaks, but may be restrictive when considering the scale of a global pandemic. For example, we may expect alarm in a rural region to be elevated prior to the first incidence, based on knowledge of disease occurrence in a nearby urban area. While one could consider adding a parameter to estimate a baseline alarm level, without strong priors this would be unidentifiable with the baseline transmissibility $\beta$ in a single population model. However, in a multi-region spatial model with some global baseline transmissibility this type of ``nugget effect" could be estimated for each region.

The proposed framework models behavioral change implicitly by allowing transmission to depend only on the perceived amount of disease in the population. However, time-varying measures of population behavior (e.g., Google mobility reports) exist and could be used to incorporate behavioral change directly in the specification of the transmission rate \citep{vanni2021use, hu2021human}. One disadvantage of this approach is it requires simulation of the behavioral change metric, which may be challenging. In contrast, transmission in the BC model is based on previous incidence, which is generated directly from the SIR model. Despite this issue, comparing an approach using mobility metrics to the BC model is a promising avenue for future work. Developments including \textit{both} measures of mobility and an alarm function based on epidemic trajectory in the transmission rate would be particularly interesting. In this setting, one could examine whether the alarm function is able to detect ``finer grain" behavioral change which may not be captured by mobility data, e.g., voluntary masking.

\section*{Acknowledgments}
Funding for the project was provided by the Canadian Statistical Sciences Institute (CANSSI) Distinguished Postdoctoral Fellowship. This research was enabled in part by computational resources provided by the University of Calgary.

\section*{Software}
Software in the form of R code, together with data used and complete documentation is available at \url{https://github.com/ceward18/epidemicBCM}.

\bibliographystyle{unsrt}  
\bibliography{references}

\end{document}